\documentclass{article}
\title{Degenerate space-time paths and the non-locality of
quantum mechanics in a Clifford substructure of space-time} 
\author{Kaare Borchsenius 
\thanks{Bollerisvej 8, 3782 Klemensker, Denmark, e-mail:
bdge@post5.tele.dk}} 
\date{19 June 2000} 
\begin{document} 
\maketitle 
\begin{abstract} The quantized canonical space-time
coordinates of a relativistic point particle are expressed
in terms of the elements of a complex Clifford algebra which
combines the complex properties of $SL(2.C)$ and quantum
mechanics. When the quantum measurement principle is adapted
to the generating space of the Clifford algebra we find
that the transition probabilities for twofold degenerate
paths in space-time equal the transition amplitudes for the
underlying paths in Clifford space. This property is used
to show that the apparent non-locality of quantum mechanics
in a double slit experiment and in an EPR type of
measurement is resolved when analyzed in terms of the full
paths in the underlying Clifford space. We comment on the
relationship of this model to the time symmetric formulation
of quantum mechanics and to the Wheeler-Feynman model.
\end{abstract}
\section{Substructure of the canonical 
space-time\\coordinates}
The fact that half-integer spin representations of the
Lorentz group are realized in nature casts doubt on the
assumption that space-time is a primary space. More
specifically, as pointed out by Penrose~\cite{penrose}, the
fact that different spatial directions of a spin-one-half
particle correspond to different complex linear combinations
of the two quantum states suggests that there is a direct
connection between the structure of space and the need for
complex state vectors in quantum mechanics. Taken together,
considerations like these point to the existence of a
substructure of space-time which combines the complex
properties of the Lorentz group and quantum mechanics.
Substructures of space-time have been discussed in Schwartz
and Van Nieuwenhuizen~\cite{schwartz} and in
Borchsenius~\cite{borch1,borch2,borch3}.

To determine the nature of such a complex substructure of
space-time we shall use the canonical quantization of a
relativistic point particle as a model. We shall adopt
Dirac's method in which space and time are treated on an
equal footing, both being regarded as functions of a
parameter-time $\tau$. Reparametrization invariance imposes
a constraint which can be used to define a Hamiltonian
together with a set of canonical variables. The quantization
results in a set of hermitian canonical space-time
coordinates, the components of which satisfy 
\begin{equation} 
X^{\mu *}_{ab}(\tau) = X^{\mu}_{ba}(\tau) 
\label{a1} 
\end{equation} 
These components transform under a Lorentz transformation in
the index $\mu$ and under a unitary change of basis in
Hilbert space in the indices $a$ and $b$. To bring out the
complex properties of the Lorentz group, we make use of the
connection between a real four-vector and a second-rank
hermitian spinor 
\begin{equation} 
V^{\mu} = \frac{1}{2}\sigma _{A\dot{B}}^{\mu}
V^{A\dot{B}}, \ \ \ \ \ \  V^{A\dot{B}} = \sigma
_{\mu}^{A\dot{B}} V^{\mu} 
\label{a2} 
\end{equation} 
where $\sigma_{\mu}$ are the four hermitian Pauli matrices.
The spinor form of the canonical space-time coordinates 
\begin{equation} 
X^{A\dot{B}}_{ab} \stackrel{\mathrm{def}}{=} \sigma
_{\mu}^{A\dot{B}}X^{\mu}_{ab} 
\label{a3} 
\end{equation} 
exhibits two hermitian properties, one related to $SL(2.C)$
and the other to the unitary group in Hilbert space. To find
a substructure of $X$ corresponding to these two groups, we
observe that the components (\ref{a3}) form a hermitian
matrix in the combined indices $(A,a)$ and $(B,b)$
\begin{equation} 
\left(X^{A\dot{B}}_{ab}\right)^{*} =
X^{B\dot{A}}_{ba} 
\label{a4} 
\end{equation} 
As shown in the appendix, any hermitian matrix can be
expressed in terms of the elements of a complex Clifford
algebra according to (\ref{e8}). For the canonical
space-time coordinates (\ref{a4}) this implies that there
exists a complex Clifford algebra with elements $C^{A}_{a}$
so that 
\begin{equation} 
X^{A\dot{B}}_{ab} = \{ C^{A}_{a},C^{*\dot{B}}_{b} \}, \ \ \
\ \ \{ C^{A}_{a},C^{B}_{b} \} =  0 
\label{a5} 
\end{equation} 
The complex linear space which generates the Clifford
algebra, and to which the $C$`s belong, we shall call
Clifford space, and we shall refer to its elements as
Clifford coordinates, borrowing from space-time terminology.
To write (\ref{a5}) in abstract form we shall adopt the
following notation. The components $C^{A}_{a}$ which
transform like a right-handed two-component spinor in the
index $A$ and as a ket vector in the index $a$ shall be
written as $\stackrel{>}{C^{A}}$ where the ket on top is
used to distinguish it from a quantum operator and an
ordinary eigenvector. Likewise $C^{*\dot{B}}_{b}$ will be
written as the bra vector $\stackrel{<}{C^{\dot{B}}} =
(\stackrel{>}{C^{B}})^{\dag}$ where $\dag$ performs both the
complex involution of the Clifford algebra and the quantum
conjugation in Hilbert space. The commutator between a ket
vector $\stackrel{>}{\chi}$ and a bra vector
$\stackrel{<}{\psi}$ shall be defined as 
\begin{equation} 
\{ \stackrel{>}{\chi } ,\stackrel{<}{\psi} \}_{ab}
\stackrel{\mathrm{def}}{=}  \{ \chi_{a} , \psi _{b}
\}, \ \ \ \ \ \{
\stackrel{<}{\psi},\stackrel{>}{\chi} \}
\stackrel{\mathrm{def}}{=} \{ \psi_{a} , \chi_{a} \}
\label{a6} 
\end{equation} 
that is, we adopt the convention that the order of the
ket and bra vectors in the first term in the commutator
determines whether \emph{both} terms are direct products or
contractions. With this notation, (\ref{a5}) can be written
in the abstract form 
\begin{equation}
X^{A\dot{B}} =
\{\stackrel{>}{C^{A}},\stackrel{<}{C^{\dot{B}}}\}, \ \ \ \ \
\{\stackrel{>}{C^{A}},\stackrel{>}{C^{B}}  \} = 0
\label{a7} 
\end{equation} 
$X$ and $\stackrel{>}{C}$ can be expressed in terms of a
complete set of eigenstates $|x_{r}\rangle$ and their
eigenvalues 
\begin{eqnarray} 
X^{\mu} =  |x_{r}^{\mu}\rangle x_{r}^{\mu} \langle
x_{r}^{\mu}| 
\label{a8} 
\\ \stackrel{>}{C^{A}} =  |x_{r}^{\mu}\rangle
c^{A}_{r}, \ \ \ \ \ c^{A}_{r}
\stackrel{\mathrm{def}}{=} \langle
x_{r}^{\mu}|\stackrel{>}{C^{A}} 
\label{a9} 
\end{eqnarray} 
When these expressions are inserted into (\ref{a7}) we
obtain 
\begin{equation}
\{c^{A}_{r},c^{*\dot{B}}_{s}\} =
\delta_{rs}x^{A\dot{B}}_{s}, \ \ \ \ \
\{c^{A}_{r},c^{B}_{s}\} = 0 
\label{a10}
\end{equation} 
Hence the eigenvalues of $X$ are determined by a set of
mutually orthogonal elements $c^{A}_{r}$ of the Clifford
algebra. To make our discussion more transparent we shall
refer to these elements as `eigenvalues' and write the
eigenstates $|x_{r} \rangle$ as $|c_{r} \rangle$. By use of
(\ref{a7}) we obtain the expression for the expectation
value of $X$ in the state $|s\rangle$ 
\begin{eqnarray} 
\langle s|X^{A\dot{B}}|s\rangle
= \langle s|\{\stackrel{>}{C^{A}},\stackrel{<}{C^{\dot{B}}}
\}|s\rangle  = \{\bar{c}^{A},\bar{c}^{*\dot{B}} \} 
\label{a11} 
\\ \bar{c}^{A} \stackrel{\mathrm{def}}{=} \langle
s|\stackrel{>}{C^{A}} 
\label{a12} 
\end{eqnarray} 
(\ref{a12}) are the Clifford coordinates corresponding to
the expectation value of the space-time coordinates.
Applying (\ref{a9}) they become
\begin{equation} 
\bar{c}^{A} = \langle s|x_{r} \rangle c_{r}^{A}
\label{a13} 
\end{equation} 
The relationship of this equation to the expression
for the expectation value of the space-time coordinates 
\begin{equation} 
\bar{x}^{\mu} = |\langle s|x_{r} \rangle|^{2}
x^{\mu}_{r} 
\label{a14} 
\end{equation} 
can be described as a linear extraction of the quantum
amplitudes as a complex substructure of the probabilities,
and of the Clifford coordinates as a complex substructure of
the space-time coordinates. If, conversely, we had sought a
substructure of space-time which had the quantum amplitudes
as a linear space of weights as in (\ref{a13}), we would
have been led to something of the nature of the
orthogonality relations (\ref{a10}).

In the continuum limit $X$ has a Continuous spectrum
and in the coordinate representation (\ref{a9}) and
(\ref{a10}) become 
\begin{equation} 
\stackrel{>}{C^{A}} =  \int_{}^{}{|x\rangle
c^{A}(x)}\,dx 
\label{a15} 
\end{equation} 
\begin{equation}
\{c^{A}(x),c^{*\dot{B}}(x')\} = x^{A\dot{B}}\delta(x-x'), \
\ \ \ \ \{c^{A}(x),c^{B}(x')\} = 0 
\label{a16} 
\end{equation} 
(\ref{a16}) generates an infinite dimensional Clifford
Algebra of a type well known from the Algebra of creation
and annihilation operators for a Fermi field.

The stability of Clifford space under $SL(2.C)$ implies that
there are at least two values $c$ and $-c$ of the Clifford
coordinates,which correspond to the same space-time
coordinates $x$. The well known degeneracy of $SO(1.3)$
transformations with respect to $SL(2.C)$ transformations is
hereby extended to space-time itself. 
\section{Canonical equations}
We consider the action:
\begin{equation}
\int_{}^{}{}L(c(\tau ),\dot{c}(\tau ))\,d\tau 
\label{f2}
\end{equation} 
Since the Lagrangian is real-valued it is natural to assume
that the Clifford variables $c$ and $\dot{c}$ occur within
anticommutators. In this case the variation of $L$ can be
expressed as
\begin{equation}
\delta L = \{\frac{\partial L}{\partial c^{A} },\delta
c^{A}\}+ c.c. +
\{\frac{\partial L}{\partial \dot{c}^{A}},\delta
\dot{c}^{A}\}+ c.c.
\label{f3}
\end{equation}
which defines the derivatives with respect to $c$ and
$\dot{c}$ up to terms which anticommute with $\delta c$. The
conjugate to $c$ is defined as 
\begin{equation}
d^{*}_{A} = \frac{\partial L}{\partial\dot{c}^{A}}
\label{f4}
\end{equation}
If $\dot{c}$ can be eliminated in favour of $d^{*}$ the
Hamiltonian becomes
\begin{equation}
H(c,d) = \{\dot{c}^{A},d^{*}_{A}\}+ c.c. - L(c,\dot{c})
\label{f5}
\end{equation}
with the equations of motion
\begin{equation}
\dot{c}^{A} = \frac{\partial H}{\partial d^{*}_{A}}, \ \ \ \
\dot{d}^{*}_{A} = -\frac{\partial H}{\partial c^{A}}
\label{f6}
\end{equation} 
In case the action (\ref{f2}) has local symmetries the
Hamiltonian is found by the methods of constrained dynamics.

We shall only consider Hamiltonians which can be expressed
in the form
\begin{equation}
H(c,d) = H(x,p), \ \ \ x^{A\dot{B}} =
\{c^{A},c^{*\dot{B}}\}, \ \ \ p_{A\dot{B}} =
\{d^{*}_{A},d_{\dot{B}}\}
\label{f7}
\end{equation}
The system corresponding to the action (\ref{f2}) cannot be
quantized in the usual way through Poisson brackets because
$c$ and $d^{*}$ become vectors $\stackrel{>}{C}$ and
$\stackrel{<}{D}$ and not operators in Hilbert space.
Instead we shall determine the conditions which have to be
imposed on $\stackrel{>}{C}$ and $\stackrel{<}{D}$ in order
to obtain the usual canonical quantization of the system
(\ref{f7}) with $p$ as the momenta conjugate to $x$.
For the Hamiltonian (\ref{f7}) the equations of motion
(\ref{f6}) become
\begin{equation}
\dot{c}^{A} = \frac{\partial H}{\partial
p_{A\dot{E}}}d_{\dot{E}}, \ \ \ \dot{d}^{*}_{A} = -
c^{*\dot{E}}\frac{\partial H}{\partial x^{A\dot{E}}}
\label{f8}
\end{equation}
The quantized form of these equations will be
\begin{eqnarray}
\frac{d}{d\tau}\stackrel{>}{C^{A}} = -
\frac{1}{2i\hbar}[H,X^{A\dot{E}}]\stackrel{>}{D_{\dot{E}}},
\ \ \ \ \frac{d}{d\tau}\stackrel{<}{D_{A}} = -
\frac{1}{2i\hbar}\stackrel{<}{C^{\dot{E}}}[H,P_{A\dot{E}}]
\label{f9}                                                 
\\ 
X^{A\dot{B}} = \{ \stackrel{>}{C^{A}},
\stackrel{<}{C^{\dot{B}}}\}, \ \ \ \ P_{A\dot{B}} =
\{\stackrel{>}{D_{\dot{B}}},\stackrel{<}{D_{A}}\}
\label{f10}
\end{eqnarray}
Applying the equations of motion (\ref{f9}) to (\ref{f10})
gives
\begin{eqnarray}
\frac{d}{d\tau}X^{A\dot{B}} =
-\frac{1}{2i\hbar}[H,X^{A\dot{E}}]\{\stackrel{>}{D_{\dot{E}}
},\stackrel{<}{C^{\dot{B}}}\} -
\frac{1}{2i\hbar}\{\stackrel{>}{C^{A}},\stackrel{<}{D_{E}}\}
[H,X^{\dot{B}E}]
\nonumber
\\ 
\frac{d}{d\tau}P_{A\dot{B}} = -
\frac{1}{2i\hbar}[H,P_{E\dot{B}}]\{\stackrel{>}{C^{E}},
\stackrel{<}{D_{A}}\}-
\frac{1}{2i\hbar}\{\stackrel{>}{D_{\dot{B}}},\stackrel{<}{C
^{\dot{E}}}\}[H,P_{A\dot{E}}]
\label{f11}
\end{eqnarray}
For these equations to reduce to the usual space-time
canonical equations of motion we must impose the commutation
relations
\begin{equation}
\{\stackrel{>}{C^{A}},\stackrel{<}{D_{B}}\}
= \delta^{A}_{B} \mu(\tau )
\label{f12}
\end{equation}
where $\mu (\tau )$ is a real scalar function of $\tau$.
Then (\ref{f11}) becomes
\begin{equation}
\frac{d}{d\tau} X^{A\dot{B}} = -
\frac{\mu (\tau)}{i\hbar}[H,X^{A\dot{B}}], \ \ \
\frac{d}{d\tau}P_{A\dot{B}}
= -\frac{\mu (\tau)}{i\hbar}[H,P_{A\dot{B}}]
\label{f13}
\end{equation}
or in reparametrized form
\begin{equation}
\frac{d}{d\bar{\tau}}X^{A\dot{B}}
= -\frac{1}{i\hbar}[H,X^{A\dot{B}}],
\ \ \  \frac{d}{d\bar{\tau}}P_{A\dot{B}}
= -\frac{1}{i\hbar}[H,P_{A\dot{B}}], \ \ \ 
\frac{d\bar{\tau}}{d\tau } = \mu (\tau)
\label{f14}
\end{equation}
Though we obtain the standard space-time canonical equations
of motion, they are subject to the (as we shall see)
important restriction that the parameter $\bar{\tau}$ is
only well defined for $\mu (\tau)\neq 0$.

Normally the compatibility of the commutation relations with
the equations of motion is ensured by the Poisson
brackets. This also applies in the present case to the
space-time commutation relations
\begin{equation}
[X^{\mu },X^{\nu }] = 0, \ \ [P_{\mu },P_{\nu }] = 0, \ \
[X^{\mu},P_{\nu }] = i\hbar\delta^{\mu }_{\nu } 
\label{f15}
\end{equation} 
which are compatible with the equations of motion
(\ref{f13}). These equations, however, assume the validity
of the Clifford commutation relations (\ref{f12}) which are
not related to any poisson brackets. We shall prove the
compatibility of these commutation relations with the
equations of motion in the classical case where they reduce
to
\begin{equation}
\{c_{A},d^{*}_{B}\} = -\epsilon_{AB}\, \mu (\tau )
\label{f16}
\end{equation}
Since all skewsymmetric second rank spinors are proportional
to $\epsilon _{AB}$, (\ref{f16}) is equivalent to the
vanishing of the symmetric part of the commutator
\begin{equation}
\{c_{(A},d^{*}_{B)}\} = 0
\label{f17}
\end{equation}
The equations of motion (\ref{f6}) subject to the constraint
(\ref{f17}) can be obtained from 
\begin{equation}
\int_{}^{}{\{\dot{c}^{A},d^{*}_{A}\}+c.c. - H(c,d) +
\lambda^{AB}(\tau )\{c_{(A},d^{*}_{B)}\}+c.c. \,d\tau}, \ \
\
\lambda^{AB} = \lambda ^{BA}
\label{f18}    
\end{equation}
by independent variation of $c$ and $d$ where $\lambda^{AB}$
are six Lagrange multipliers. A local $SL(2.C)$
transformation
\begin{equation}
c_{A} = S_{A}^{\ E}(\tau ) \bar{c}_{E}, \ \ \ \ d^{*}_{A} =
S_{A}^{\ E}(\tau ) \bar{d}^{*}_{E}
\label{f19}
\end{equation}
turns (\ref{f18}) into
\begin{equation}
\int_{}^{}{\{\dot{\bar{c}}^{A},\bar{d}^{*}_{A}\}+c.c. -
H(\bar{c},\bar{d}) + (\dot{S}^{EA}S_{E}^{\ B}+
\bar{\lambda}^{AB})\{\bar{c}_{(A},\bar{d}^{*}_{B)}\}+c.c.
\,d\tau}
\label{f20}
\end{equation}
The last two terms in (\ref{f20}) can be made to vanish if
\begin{equation}
\dot{S}^{EA}(\tau)S^{\ B}_{E}(\tau) = -
\bar{\lambda}^{AB}(\tau )
\label{f21}   
\end{equation}
Taking $\lambda$ to be small, the infinitesimal $SL(2.C)$
transformation
\begin{equation}
S^{AB}(\tau ) = \epsilon ^{AB} + \kappa^{AB}(\tau ), \ \ \ \
\kappa^{AB} = \kappa^{BA}
\label{f22}
\end{equation}
turns (\ref{f21}) into
\begin{equation}
\dot{\kappa}^{AB} = -\lambda ^{AB}
\label{f23}
\end{equation}
which can always be solved for $\kappa^{AB}(\tau )$ in terms
of $\lambda ^{AB}(\tau)$. The constraint (\ref{f17}) can
therefore be absorbed into a local $SL(2.C)$ transformation
of the dynamical variables and will accordingly preserve the
form of the equations of motion. In section 4 we shall
examine a specific model of the relativistic point particle
and find that also the quantum form of the Clifford
commutation relations leads to a consistent result.
\section{Degenerate space-time paths}
When $\mu (\tau )$ in (\ref{f12}) has a zero, the parameter
$\bar{\tau }$ of the space-time equations of motion is
ill-defined. We should therefore be prepared to encounter
complete solutions $C(\tau)$,$D(\tau)$ to the equations of
motion (\ref{f11}) which generate incomplete solutions
$X(\tau)$,$P(\tau)$ to the space-time equations of motion.
To understand what happens, let us assume that $\mu (0)=0$.
Then for $\tau =0$ the commutation relations
(\ref{f12}) reduce to
\begin{equation}
\{\stackrel{>}{C^{A}}(0),\stackrel{<}{D_{B}}(0)\} = 0
\label{g1}
\end{equation}
Let us expand $C(\tau )$
\begin{equation}
\stackrel{>}{C}(\tau) = \stackrel{>}{C}(0) + \cdots
+ \frac{1}{n!}\stackrel{>}{C^{(n)}}(0)\tau ^{n} +  \cdots 
\label{g2}
\end{equation}
The higher order derivatives $C^{(n)}(\tau )$ are
obtained by differentiating the equations of motion
(\ref{f9}) and reinserting the expressions for $\dot{X}$ and
$\dot{P}$ obtained from (\ref{f11}). Because of the
commutation relations (\ref{g1}) this can only result in
coefficients which contain terms of the form 
\begin{equation}
F^{A}_{B}(X(0),P(0))\stackrel{>}{C^{B}}(0) \ \ \mbox{or} \ \
G^{A}_{\dot{B}}(X(0),P(0))\stackrel{>}{D^{\dot{B}}}(0)
\label{g3}
\end{equation}
When $C(\tau)$,$D(\tau )$ is a solution to (\ref{f9}), so
is $-C(-\tau)$,$D(\tau )$. Thus $C(\tau)$ must be odd under
a change of sign of $C(0)$ and $\tau $. It follows that in
the expansion (\ref{g2}) all terms of even order must have
coefficients of the first type in (\ref{g3}) and all terms
of odd order must have coefficients of the second type. When
therefore the expansion (\ref{g2}) is inserted into
(\ref{a5}) to determine $X(\tau )$ we find that, because of
the commutation relations (\ref{g1}), all
anti-commutators between terms of odd order and terms of
even order vanish. Accordingly $X(\tau )$ can only contain
terms of even order and must therefore be an even function
of $\tau$: 
\begin{equation}
X(-\tau ) = X(\tau )
\label{g4}
\end{equation}  
This implies that $C(\tau)$ reproduces $X(\tau)$ twice,
making it twofold degenerate. Hence there exist complete
paths in Clifford space which have either a beginning or an
end in physical time. We shall assume that it is the first
possibility which applies, and to avoid any contradiction
with experience we must assume that the starting time lies
so far back as to put it under the provision of cosmology.
For all particles to have the same starting time we must
assume that they are all states of more fundamental objects
to which the Clifford substructure in some form can be
applied.

The classical paths will, like $X(\tau )$, be even functions
of $\tau$. In the quantum regime, however, paths for which
$x(\tau)\neq x(-\tau )$ will also contribute to the
transition amplitudes. Consequently the Clifford model will
seem to be non-local from a space-time point of view. We
shall interpret this non-locality in section 6.
\section{The relativistic point particle}
Since there exists no $SL(2.C)$ invariant hermitean second
rank spinor, but only the real skewsymmetric metric
$\epsilon _{AB}$, the simplest reparametrization invariant
action for a relativistic point particle which only depends
on $\dot{c}$ is 
\begin{equation}
-2^{\frac{7}{4}}\sqrt{m}\int_{}^{}{\sqrt[4]{\{\dot{c}^{A},
\dot{c}^{*\dot{B}}\}\{\dot{c}_{A},\dot{c}^{*}_{\dot{B}}\}}\,
d\tau }
\label{h1}
\end{equation}
The conjugate to $c$ is
\begin{equation}
d^{*}_{A} = -2^{\frac{3}{4}}\sqrt{m}(\{\dot{c}^{E},
\dot{c}^{*\dot{F}}\}\{\dot{c}_{E},\dot{c}^{*}_{\dot{F}}\})
^{-
\frac{3}{4}}\{\dot{c}_{A},\dot{c}^{*}_{\dot{B}}\}\dot{c}^{*
\dot{B}}
\label{h2}
\end{equation}
Not unexpectedly the Hamiltonian (\ref{f5}) vanishes
because of reparametrization invariance. By use of the
relation
\begin{equation}
V_{A\dot{F}}V^{B\dot{F}} =
\delta ^{B}_{A} V_{\mu }V^{\mu }
\label{h3}
\end{equation}
for a hermitean second rank spinor, we obtain from
(\ref{h2}) the associated constraint
\begin{equation}
\frac{1}{2}\{d^{*}_{A},d_{\dot{B}}\}\{d^{*A},d^{\dot{B}}\} =
m^{2}
\label{h4}
\end{equation}
or
\begin{equation}
p_{\mu }p^{\mu } = m^{2}
\label{h5}
\end{equation}
This is the same constraint as would have been obtained from
the usual space-time Lagrangian $m\sqrt{\dot{x}^{2}}$, but
with the important difference that $p_{\mu }$ is no longer a
primary dynamical variable. The new Hamiltonian is
proportional to the constraint:
\begin{equation}
H(\stackrel{>}{C},\stackrel{<}{D}) = \nu (\tau )(P_{\mu
}P^{\mu }-m^{2})
\label{h6}
\end{equation}
The gauge is fixed by choosing $\nu (\tau ) = \frac{1}{2m}$.
By use of the space-time commutation relations the
equations of motion (\ref{f9}) become
\begin{equation}
\frac{d}{d\tau}\stackrel{>}{C^{A}} =
\frac{1}{2m}P^{A\dot{E}}\stackrel{>}{D_{\dot{E}}}, \ \ \ \ \
\frac{d}{d\tau}\stackrel{<}{D_{A}} = 0
\label{h7}
\end{equation}
with the solution
\begin{equation}
\stackrel{>}{C^{A}}(\tau ) =
\stackrel{>}{C^{A}}(0)+\frac{1}{2m}P^{A\dot{E}}(0)\stackrel{
>}{D_{\dot{E}}}(0)\tau , \ \ \ \ \stackrel{<}{D_{A}}(\tau )
= \stackrel{<}{D_{A}}(0)
\label{h8}
\end{equation}
From (\ref{h8}) we obtain   
\begin{equation}
\{\stackrel{>}{C^{A}}(\tau ),\stackrel{<}{D_{B}}(\tau )\} =
\{\stackrel{>}{C^{A}}(0),\stackrel{<}{D_{B}}(0)\}+\frac{1
}{2m}P_{B\dot{E}}(0)P^{A\dot{E}}(0) \tau 
\label{h9}
\end{equation}
Applying (\ref{h3}) and the quantum form of (\ref{h5}) to
(\ref{h9}) it becomes
\begin{equation}
\{\stackrel{>}{C^{A}}(\tau ),\stackrel{<}{D_{B}}(\tau )\} =
\{\stackrel{>}{C^{A}}(0),\stackrel{<}{D_{B}}(0)\}+ \delta
^{A}_{B} \frac{m}{2}\tau 
\label{h10}
\end{equation}
Accordingly, the Clifford commutation relations (\ref{f12})
are preserved in time by the equations of motion, and with
the choice $\tau = 0$ for the zero-point of $\mu(\tau)$ we
obtain   
\begin{equation}
\mu (\tau ) = \frac{m}{2}\tau , \ \ \ \ \bar{\tau } =
\frac{m}{4}\tau
^{2}
\label{h11}
\end{equation}
The corresponding space-time solution is
\begin{equation}
X^{\mu }(\bar{\tau } ) = X^{\mu }(0) + \frac{1}{m}P^{\mu
}(0) \bar{\tau }, \ \ \ \ P_{\mu }(\tau ) = P_{\mu }(0 )
\label{h12}
\end{equation}
In accordance with the general result in section 3, the
complete solutions to the Clifford equations of motion
(\ref{h7}) are double coverings of the incomplete solutions
$X(\bar{\tau}), \bar{\tau }\geq 0$ to the space-time
equations of motion.
\section{Measurement principle}  
The measurement principle in quantum mechanics says
that the (abstract) state vector is constant in time
as long as no measurement is being performed. After
a measurement has been performed the state vector is
replaced by the eigenvector of the measured quantity
for subsequent times (`state vector reduction').
This measurement principle applies equally well to Dirac's
parameter-time formalism when `time' is taken to be a
parameter-time with the same direction as our $\bar{\tau }$
in the foregoing. Recognizing the primary character of
Clifford space, we shall instead assume that the reduction
of the state vector takes place in the positive direction of
$\tau$ itself which therefore comes to represent the true
direction of causality: 
\begin{quote} 
Measurement Principle. \emph{The state vector of the
particle is constant in parameter-time $\tau$ as long as no
measurement is being performed. When the particle is
measured to be in the eigenstate $|x_{P}\rangle$ of $X$ the
state vector is replaced by $|c_{P}\rangle = |x_{P}\rangle$
for parameter-times $\tau > \tau_{P}$ where $c_{P}$ is an
`eigenvalue' of $\stackrel{>}{C}(\tau_{P})$ and
$c_{P}$ and $\stackrel{>}{C}(\tau_{P})$ satisfy
$\{c_{P},c_{P}^{*}\} = x_{P}$ and
$\{\stackrel{>}{C}(\tau_{P}),
\stackrel{<}{C}(\tau_{P})\} = X$ respectively}
\end{quote} 
Using a convenient terminology we shall say that the
Clifford position of the particle has been measured to be
$c_{P}$ at $\tau = \tau_{P}$. The measurement principle
respects the fact that since the interaction-Hamiltonians
used for measuring space-time positions depend only on
$C$ \emph{through} $X$, the state vector reduction in
Clifford space should also be defined trough $X$ and its
eigenvalues. In the following we shall examine the
consequences of this principle. 

Let the space-time position of the particle have been
measured to be $x_{Q}$. From (\ref{g4}) and (\ref{a10}) it
follows that $C(\tau)$ satisfies the criteria in the
measurement principle at two parameter-times $\tau =
\pm \tau_{Q}$. Let us call the corresponding `eigenvalues'
for $c_{Q+}$ and $c_{Q-}$. Hence the state vector will be
$|c_{Q- }\rangle$ and $|c_{Q+}\rangle$ (both equal to
$|x_{Q}\rangle$ ) right after $\tau = -\tau_{Q}$ and
$\tau = \tau_{Q}$ respectively. If no measurement is
being performed between $\tau = -\tau_{Q}$ and $\tau
= \tau_{Q}$ the particle will arrive at $\tau =
\tau_{Q}$ in the state $|c_{Q-}\rangle$. Since after $\tau =
\tau_{Q}$ the state is $|c_{Q+}\rangle$, the transition
amplitude is $\langle c_{Q-}|c_{Q+}\rangle = \langle
x_{Q}|x_{Q}\rangle = 1$. Therefore the measurement
principle is self-consistent as long as no measurement is
being performed between $\tau = -\tau_{Q}$ and $\tau =
\tau_{Q}$. Let us now assume that such a measurement
\emph{is} being performed, resulting in the space-time
position $x_{P}$ corresponding to the Clifford positions $
c_{P\pm}$ at $\tau = \pm \tau_{P}$ respectively, where
$\tau_{P} < \tau_{Q}$. The transition amplitude for
the particle to pass from $c_{Q-}$ through  $c_{P-}$
and $c_{P+}$ to $c_{Q+}$ is 
\begin{equation} 
\langle c_{Q-}| c_{P-}\rangle \langle c_{P+}|
c_{Q+}\rangle = |\langle x_{P}| x_{Q}\rangle|^{2}
\label{b16} 
\end{equation} 
and therefore equals the transition probability for the
particle to move from $x_{P}$ to $x_{Q}$. We conclude that
the space-time transition probabilities arise as transition
amplitudes for the complete paths in Clifford space.

Note that viewed from space-time it appears as if there are
two amplitudes, one moving forward in time from $x_{P}$ to
$x_{Q}$ and the other moving backwards in time from $x_{Q}$
to $x_{P}$. This resembles the situation in the time
symmetric formulation of quantum mechanics by Aharonov and
Vaidman~\cite{ahavai}, Costa de Beauregard~\cite{costa}, and
Werbos~\cite{werbos}. In the present model the two state
vectors of time symmetric quantum mechanics are recognized
to be one and the same, propagating along a path which
covers the space-time path twice. The use of parameter-time
in our model is necessitated by the secondary character of
physical time, but it has the added advantage of ensuring
manifest Lorentz invariance. 

The present model should also be compared to the so-called
`double space-time interpretation of quantum mechanics'
Bialynicki-Birula~\cite{birula}, inspired by Schwinger's
time loop integrated amplitudes. The main problem in this
interpretation is how to join the two space-time sheets at
infinity to allow a particle to travel along a single path
on the two sheets.
 
The choice of taking the causal direction of state vector
reduction to be in the positive direction of parameter-time
$\tau$ rather than of `affine time' $\bar{\tau} $ strongly
suggests that the same should apply to the direction of
propagation of classical fields. The following heuristic
observation shows that this is not necessarily inconsistent
with experience. Let the union of all possible particle
trajectories for $\tau \leq 0$ and for $\tau \geq 0$ form
regions $\Omega_{-}$ and $\Omega_{+}$ of Clifford space
which correspond to the same space-time region. For the
field to propagate in the positive direction of $\tau$ we
should choose the advanced field on $\Omega_{-}$ and the
retarded field on $\Omega_{+}$. The contribution to the
electrodynamic action in the proper-time interval
$[\bar{\tau}_{1};\bar{\tau}_{2}]$ of a test-particle with
charge $e$ traversing this region is
\begin{eqnarray} 
\frac{1}{2}\int_{-\tau_{2}}^{-\tau_{1}
}{m\sqrt{\dot{x}^{2}} + A_{adv}e(-\dot{x})\,d\tau }
+ \frac{1}{2}\int_{\tau_{1}}^{\tau_{2}}{m\sqrt{
\dot{x}^{2}} + A_{ret}e\dot{x}\,d\tau} \nonumber\\
= \int_{\bar{\tau}_{1}}^{\bar{\tau}_{2}}{m\sqrt{\dot{x}
^{2}} + (\frac{1}{2}A_{adv} +
\frac{1}{2}A_{ret})e\dot{x}\,d\bar{\tau}} 
\label{d1} 
\end{eqnarray} 
The test-particle will therefore detect the effective field
to be the time symmetric half-advanced plus half-retarded
field. Assuming complete absorption and no self-interaction
Wheeler and Feynman~\cite{whefey} have shown that this time
symmetric field leads to the conventional rules of
electrodynamics.
\section{Interpretation of non-locality}
Consider a particle which travels in space-time from a point
$P$ to a point $Q$ and is forced to travel trough two
alternative points $S_{1}$ and $S_{2}$. This corresponds to
the double slit experiment with the two slits being opened
at given times. As follows from our foregoing discussion the
particle can follow four alternative sets of paths in
Clifford space corresponding to the the four sequences of
positions in Clifford space ordered according to parameter
time: 
\begin{equation} 
c_{Q-}, \  c_{S_{i}-}, \  c_{P-}, \  c_{P+}, \
c_{S_{j}+}, \  c_{Q+}, \  i,j = 1,2 
\label{c1} 
\end{equation} 
The amplitude for the particle to travel from $c_{Q-}$ to
$c_{Q+}$ is the sum of the amplitudes for all four different
sets of paths 
\begin{eqnarray}
 \sum_{i,j=1}^{2}{\langle c_{Q-}|c_{S_{i}-}\rangle \langle
c_{S_{i}-}|c_{P-}\rangle\langle
c_{P+}|c_{S_{j}+}\rangle\langle
c_{S_{j}+}|c_{Q+}\rangle}\nonumber \\ = |\langle
x_{P}|x_{S_{1}}\rangle\langle x_{S_{1}}|
x_{Q}\rangle + \langle x_{P}|x_{S_{2}}\rangle\langle
x_{S_{2}}|x_{Q}\rangle|^{2} 
\label{c2} 
\end{eqnarray} 
which is the well known probability for the particle to
travel from $P$ to $Q$. The customary interpretation of this
transition probability is that there are two alternative
paths and that the transition probability is the sum of the
probabilities for each path plus two interference terms
which seem to signal a non-local influence of one path on
the other. From (\ref{c2}) we see that there are really four
different sets of paths and that the two interference terms
are the amplitudes for the two sets of paths where the
particle goes through each slit at opposite parameter times.
The apparent non-locality can be entirely attributed to
the twofold degeneracy of the space-time paths. 

If we measure the position of the particle at one of the
slits, say $S_{1}$, then according to our measurement
principle the particle has to travel through both $c_{S_{1}-
}$ and $c_{S_{1}+}$ or neither of them, and this excludes
the two sets of paths where the particle passes through both
slits. This removes the interference terms in accordance
with the space-time view of quantum mechanics. This analysis
is readily extended to a many-slit experiment by observing
that all interference terms arise from pairs of slits.

As the second example of non-locality we shall consider an
EPR type of measurement. Consider a composite system $(PQ)$
consisting of two spin $\frac{1}{2}$ particles $P$ and $Q$.
First the position and the total spin of the composite
system is measured to be $x_{(PQ)}$ and 0. After this
measurement $P$ and $Q$ become separated by a spacelike
distance and their position and spin along some axis are
measured to be $x_{P}$ and $\frac{1}{2}$ and $x_{Q}$ and $-
\frac{1}{2}$ respectively. The last two measurements appear
to be correlated despite the spacelike separation of $P$
and $Q$, giving thereby  the impression of `action at a
distance'. However, according to our measurement principle,
the position measurements, and together with them the spin
measurements, each correspond to two measurements in
Clifford space at opposite values of $\tau$ . If the
measurements of $x_{(PQ)}$, $x_{P}$ and $x_{Q}$ correspond
to parameter-times $\tau = \pm \tau _{(PQ)}$, $\tau =
\pm \tau _{P}$ and $\tau = \pm \tau _{Q}$ respectively, then
the sequence of events for negative $\tau$ can be described
as follows. First at parameter-times $\tau = -\tau_{P} $
and $\tau = -\tau_{Q} $ the spins along some axis of $P$ and
$Q$ are measured to be $\frac{1}{2}$ and $-\frac{1}{2}$
respectively. At the later parameter-time $\tau = -\tau
_{PQ} > -\tau_{P}, -\tau_{Q}$, $P$ and $Q$ merge into a
composite system $(PQ)$ and the total spin is  measured to
be 0 . We would not object to this last sequence of events
because it suggests no correlation between the spacelike
separated states of $P$ and $Q$. Rather, it suggests an
obvious correlation between the states of $P$ and $Q$ on the
one hand and the state of the composite system $(PQ)$ on
the other, which invokes no need for `action at a distance'.
Nevertheless these two sequences of events, corresponding
to opposite values of $\tau$, together form a single series
of events in the causal direction of state vector reduction
in Clifford space and are both the result of the same
space-time measurements on a degenerate space-time path.
They are therefore on an equal footing and we conclude that
it has no absolute meaning to say whether the
spin-measurements on $P$ and $Q$ are correlated or
independent. Accordingly, the apparent manifestation of
`action at a distance' loses its significance.
\appendix \section{Appendix. Clifford algebras and
Hermitian quadratic forms}
A real Clifford algebra arises naturally as the `square
root' of a real quadratic form $Q$ on a linear space $V$: 
\begin{equation} 
v^{2}=Q(v) , \  v \in  V 
\label{e1} 
\end{equation} 
$Q$ can have any signature $(N_{+},N_{0},N_{-})$ .
In case $Q$ is degenerate ($N_{0}\neq 0$), the algebra
contains Grassmann elements. When $v$ is expanded on an
orthogonal basis $e_{i}$ of $V$ it follows that (\ref{e1})
is satisfied if
\begin{equation} 
\frac{1}{2}\{e_{i},e_{j}\} = \delta _{ij}Q(e_{i})
\label{e2} 
\end{equation} 
The basis $e_{i}$ generates the Clifford algebra. Consider
now a quadratic form $Q$ with signature 
$(2N_{+},2N_{0},2N_{-})$. We can rearrange the generators
$e_{i}$ into two sets $a_{i}$ and $b_{i}$, $i=1, \ldots ,N$
which when normalized satisfy 
\begin{equation} 
\frac{1}{2}\{a_{i},a_{j}\} =
\frac{1}{2}\{b_{i},b_{j}\} = \delta_{ij}Q(a_{i}), \
\ \ \ \ \{a_{i},b_{j}\} = 0 
\label{e3} 
\end{equation} $a_{i}$ and $b_{i}$ can be used as `real' and
`imaginary' parts to define the complex quantities 
\begin{equation} 
f_{j} = a_{j} + i \, b_{j} 
\label{e4} 
\end{equation} 
where $i$ is the imaginary unit. The elements $f_{j}$ are
seen to satisfy the commutation relations 
\begin{equation} 
\frac{1}{4}\{ f_{i},f_{j}^{*} \} = \delta
_{ij}Q(a_{i}), \ \ \ \ \ \{f_{i},f_{j}\} = 0
\label{e5} 
\end{equation} 
where $*$ is any complex involution induced by a
self-involution in the real algebra. The algebra generated
by $f_{i}$ is a complex Clifford algebra.

For any  hermitian quadratic form $H$ on a linear space $V$
there exists a complex Clifford algebra generated by $V$
which satisfies 
\begin{equation} 
\frac{1}{2}\{v,v^{*}\} = H(v) , \  v \in V, \ \ \ \
\ v^{2} = 0 , \  v \in  V 
\label{e6} 
\end{equation} 
The proof follows by expanding $v$ on an orthogonal basis
$f_{i}$ of $V$. (\ref{e6}) is seen to be satisfied if 
\begin{equation} 
\frac{1}{2}\{ f_{i},f_{j}^{*} \} = \delta
_{ij}H(f_{i}), \ \ \ \ \ \{ f_{i},f_{j} \} = 0
\label{e7} 
\end{equation} 
which is recognized to be the generating algebra of a
complex Clifford algebra. Expressed in matrix language,
(\ref{e6}) implies that any hermitian matrix $H_{ij}$ can be
expressed in terms of elements $v_{i}$ of a complex
Clifford
algebra:
\begin{equation} 
H_{ij} = \{v_{i},v_{j}^{*}\}, \ \ \ \ \
\{v_{i},v_{j}\} = 0 
\label{e8} 
\end{equation}
 
\end{document}